# Tunable bandgaps and flat bands in twisted bilayer biphenylene carbon

Yabin Ma(马亚斌)[1,2], Tao Ouyang(欧阳滔)[1], Yuanping Chen(陈元平)[1,2*], Yuee Xie(谢月娥)[1,2*]

[1]*School of Physics and Optoelectronics, Xiangtan University, Xiangtan, Hunan, 411105, China*

[2]*Faculty of Science, Jiangsu University, Zhenjiang, 212013, Jiangsu, China*

**Abstract**

Owing to the interaction between the layers, the twisted bilayer two-dimensional (2D) materials exhibit numerous unique optical and electronic properties different from the monolayer counterpart, and have attracted tremendous interests in current physical research community. By means of first-principles and tight-binding model calculations, the electronic properties of twisted bilayer biphenylene carbon (BPC) are systematically investigated in this paper. The results indicate that the effect of twist will not only leads to a phase transition from semiconductor to metal, but also an adjustable band gap in BPC (0 meV to 120 meV depending on the twist angle). Moreover, unlike the twisted bilayer graphene (TBG), the flat bands in twisted BPC are no longer restricted by "magic angles", i.e., abnormal flat bands could be appeared as well at several specific large angles in addition to the small angles. The charge density of these flat bands possesses different local modes, indicating them might be derived from different stacked modes and host different properties. The exotic physical properties presented in this work foreshow twisted BPC a promising material for the application of terahertz and infrared photodetectors and the exploration of strong correlation.

*Correponding author: chenyp@ujs.edu.cn and yueex@ujs.edu.cn.

Keywords: twisted bilayer biphenylene carbon; adjustable band gap; flat bands

# 1. Introduction

   The rise of graphene has brought the field of 2D materials to the center stage of condensed matter physics and materials science[1-4]. Recently, the attention has been focused on the bilayer graphene (BLG)[5, 6], especially the twisted bilayer graphene (TBG)[7-10]. Although the weak van der Waals force dominate the inter-layer interaction in TBG, it still greatly affects the interlayer electron motion due to the lattice deformation, which gives rise to the suppressed band dispersion at the "magic angle" and the Coulomb interaction dominated electronic behaviors[11-14]. Many novel physical properties have also been found in the TBG, such as Mott-like insulator at half-filling, unconventional superconductivity similar to the high temperature superconductors[15-17], anomalous quantum Hall effect at the "magic angle"[18, 19], and orbital magnetism caused by the topological flat band[20-22]. These exotic physical properties make the TBG a reliable material platform for the investigation of strong correlation and designing quantum computers. In addition to the TBG, there are many investigations about other twisted 2D materials. For instance, flat bands could also be observed in twisted bilayer boron nitride, and such flat bands not only strongly modify the optical properties and structure of excitons, but also support topological superconductivity upon hole (or electron) doping[23-25]. Similar phenomenon is revealed in twisted transition metal dichalcogenides (TMDs) as well[26-29]. Up to data, the previous studies on the twisted 2D materials are mainly concerned with the zero bandgap or wide bandgap materials while the exploration about the narrow bandgap materials is still quite limited. However, narrow bandgap materials are very important for the study of light absorption and photoelectric detection in terahertz and infrared wavelengths，and it remains a daunting challenge to search for the materials that possess suitable bandgaps in corresponding wavelength range. So that the twisted narrow bandgap materials are worth for more and deepen study, which also provide a new perspective for the understanding of mechanism of the unique electronic properties of twisted 2D materials.

The diversity of hybrid forms endows element carbon the ability for designing a variety of 2D allotropes with novel physical properties[30-33]. In addition to graphene. numerous 2D carbon allotropes have been proposed and investigated, e.g., α-, β-, γ-R-, 6, 6, 12-graphyne[34-36], Stone-Wales graphene (SW40), penta-graphene[37], and Kagome graphene. These carbon sheet not only share excellent thermal and chemical stability, but also exhibit rich fascinating electronic properties, for example, intrinsic type-III Dirac cone and excellent stability (just about 133 meV higher than graphene) in the SW40[38], Dirac-like fermions in the R-graphyne nanoribbons[39] and Wigner crystallization in the Kagome graphene with a partial filling factor of the flat bands[40]. Among these 2D carbon materials, biphenylene carbon (BPC) is a semiconductor with a narrow band gap, and possesses great potential applications in high performance electronic devices and efficiency hydrogen separation[41-43]. More interestingly, based on 1,3,5-trihydroxybenzene through dehydration and polymerization reactions, Du et al. have successfully synthesized BPC nanosheets in experiments[44]. In view of the intrinsic narrow band gap and experimental breakthrough, it is natural to consider whether there will exist some unusual phenomena in the twisted BPC, and thus broadening the field of twisted 2D materials.

Inspired by this, we systematically investigate the electronic properties of twisted BPC in this work. The calculations show that the bandgap is dependent on the twist angle, and a transition from semiconductor to metal is identified for twisted BPC. In contrast to TBG, beyond the 'magic angles' the unique flat bands in twisted BPC could occurs as well for several particular large angles. These flat bands are separated from other bands and their bandwidth is all less than 14 meV. Through analyzing the charge density, we find that these flat bands are accompanied by different local modes. The rest of this paper is organized as follows. In the next section, we give a brief description of the computation details about the tight-binding model in this work. In Sec. III, the electronic band structure of twisted BPC with AA and AB stacking are presented. The charge density of the flat bands is also analyzed. Finally, our conclusions are summarized in Sec. IV.

## 2. Models and methods

The monolayer BPC structure could be seen as replacing each atom in graphene with a carbon six-membered ring, the unit cell contains 12 carbon atoms. After full optimization, the lattice parameter of the monolayer BPC is a = b = 6.767 Å. There are three different bond lengths in the monolayer BPC are 1.366, 1.47 and 1.48 Å, respectively. Based on the different stacked methods (AA-stacked in Fig. 1(a) and AB-stacked in (Fig. 5(a)). we consider two twisted modes (AA-twisted and AB-twisted), as shown in Fig. 1(b) and Fig. 5(b). In both cases, the top green layer is twisted with respect to the bottom purple layer. The rotation axis of AA-twisted is along the z axis at the center of the twelfth-membered ring. The rotation axis of AB-twisted is along the z axis at the common center of the twelfth-membered ring at the bottom and the fourth-membered ring at the top. To characterize the atomic coordinates of the twisted structures, we define a set of basis vectors $\boldsymbol{a_0}$ and $\boldsymbol{b_0}$ for monolayer BPC, defined as $\boldsymbol{a_0} = \frac{\sqrt{3}}{2}a\boldsymbol{i}$ and $\boldsymbol{b_0} = \frac{1}{2}b\boldsymbol{j}$, where $\boldsymbol{i}$ and $\boldsymbol{j}$ are the unit vectors along *x* and *y*, respectively (Fig. 1(a)). To describe the second layer BPC, one may use a rotation matrix $R[\theta] = \begin{bmatrix} cos\theta & -sin\theta \\ sin\theta & cos\theta \end{bmatrix}$, instead, where $\theta$ is the twist angle.

The twist angle $\theta$ is related to $(m, n)$ by

$$\theta = 2\, tan^{-1}\left(\frac{n}{m}\right)\left(\frac{b_0}{a_0}\right). \tag{1}$$

As such, the lattice constant of the twisted BPC is related to $(m,n)$ by $A = \binom{m}{n}$, $B = RA$. Due to the $C_6$ symmetry, one may confine angle $\theta$ in the range of 0° to 30°.

The tight-binding model for $p_z$ atomic orbitals is used to model calculation. The Hamiltonian is written as:

$$H = H_1 + H_2, \tag{2}$$

$$H_1 = \sum_{<p,q>}\sum_\mu t_{pq}\, e^{-i\boldsymbol{k}\cdot\boldsymbol{d}_{pq}^\mu}, \tag{3}$$

$$H_2 = \sum_{<p,p'>}\sum_v t_{pp'} e^{-ik\cdot d^v_{pp'}} + \sum_{<q,q'>}\sum_w t_{qq'} e^{-ik\cdot d^w_{qq'}}. \tag{4}$$

$H_1$ is the interlayer interaction and $H_2$ is the Hamiltonian for the two monolayers. Indices $p$ and $q$ represent the atoms in the two layers, $d^\mu_{pq}$, $d^v_{pp'}$ and $d^w_{qq'}$ are the relative position vectors between atomic pairs: $(p,q)$, $(p,p')$, and $(q,q')$, and $t_{pq}$, $t_{pp'}$ and $t_{qq'}$ are the corresponding hopping energies, which are defined according to the Slat-Koster semi-empirical formula[45-47], e.g.:

$$-t_{pq} = V_{pp\pi}\left[1 - \left(\frac{d_0}{d_{pq)}}\right)^2\right] + V_{pp\sigma}\left(\frac{d_0}{d_{pq}}\right)^2, \tag{5}$$

$$V_{pp\pi} = V^0_{pp\pi}\exp(-\frac{d_{pq}-c_0}{\delta}), \tag{6}$$

$$V_{pp\sigma} = V^0_{pp\sigma}\exp(-\frac{d_{pq}-d_0}{\delta}), \tag{7}$$

where $d_0$ (= 3.30 Å) is the vertical distance between the layers, $c_0$ = 1.35 Å is the nearest neighbor C-C bond length, $d_{pq}$ is the interatomic distance, and $\delta$ is a decay length, we choose $\delta$ to be 0.184 $a$ ($a$ = 5.20 Å), corresponding to having a next-nearest intralayer coupling = $0.1 V^0_{pp\pi}$ [48]. $V^0_{pp\pi}$ is the intralayer transfer integral, $V^0_{pp\sigma}$ is the interlayer transfer integral. For $t_{pp'}$ and $t_{qq'}$, only the three nearest neighboring interactions $t_1, t_2$ and $t_3$ shown in Fig. 1(a), are considered. By simulating the DFT band structures of monolayer BPC, AA-stacked, AB-stacked and twisted 21.79° (see Fig. S2), we arrive at the above TB parameters. For example, $t_1$= -2.65 eV, $t_2 = t_3 =$ -2.56 eV, $V^0_{pp\pi}$ = -2.65 eV, $V^0_{pp\sigma}$ = 0.38 eV. The results of tight-binding model agree well with the first-principles calculations.

3. **Result and discussion**

Before studying the twist bilayer BPC, we calculated the band structure of monolayer BPC. From Fig. S2(a), one can clearly see that monolayer BPC is a semiconductor with a direct narrow band gap (42 meV), and the conduction band minimum (CBM) and valence band maximum (VBM) both located at the K point, which agrees quite well with the previous literature[49]. For bilayer BPC, there exists two different stacked types, i.e., AA-stacked and AB-stacked. We firstly calculate the effect of twist angle on the electronic structure of AA-stacked BPC. The results show that the AA-stacked BPC (without twisted, twist angle is 0°) is a metal, and its detailed electronic band structure is depicted in Fig. 2(a). When the twist angle is taken into account, the AA-stacked BPC transforms from metal to semiconductor, in which only the structure with twisted angel 21.79° is indirect bandgap semiconductor (shown in Fig. 3(c)). This result indicates that owing to the interlayer interaction there exists obvious phase transition in the AA-twisted bilayer BPC.

In order to show the influence of twist angle on the band gap more clearly, in Fig. 3 we plot the variation of band gap of AA-twisted BPC with the twist angle. One can see that the variation of band gap with twist angle could be divided into four regions. For region I, the band gap presents a constantly downshift with the increasing twist angle, and it decreases to about 25 meV with twisted angel 5.09°. For region II, the band gap first increases rapidly, and then slowly as the twist angle continues to increase, and gradually reaches the maximum value in this region (about 100 meV with twisted angle 13.17°, and the detailed band structure is shwon in Fig. 2(b)). When the twist angle enters to region III, the band gap sharply declines with the increasing twist angle and reduces to the minimum value of the entire twisted system when the twist angle becomes larger until 21.79° (about 8 meV), the corresponding band structure is depicted in Fig. 2(c). For region IV, the band gap possesses a sharp upshift as twist angle gets larger. When the twist angle increases to 29.4°, the band gap reaches the maximum value of the entire twisted system (120 meV), which is nearly three times that of monolayer BPC and its band structure is given in Fig. 2(d). The results illustrated in Figs 2 and 3 indicate that the band gap of bilayer BPC could be effectively engineered through the twist angle. Moreover, it is worth to mention that the controllable bandgap of twisted bilayer BPC offers the intriguing opportunity to design optoelectronic detection systems. The bandgap is in the region of terahertz and infrared radiation, implying one can use such twisted system to design and fabricate corresponding photodetectors.

It is well known that the band structure of TBG will undergoes a transition from Dirac band to flat band structure, and its physical phase changes from topological semimetal to Mott insulator or superconductor accordingly. However, it should be pointed out that such phenomenon occurs only in the case of "magic angles". Our calculations show that the unique flat-band structures are appeared as well in the twisted bilayer BPC. Similar to TBG, when the twist angle is less than 6.0°, two flat bands are respectively generated at VBM and CBM of AA-twisted BPC, and these flat bands are energetically separated (as shown in Fig. 4(a-b)). In contrast to the TBG, the flat bands are also emerged at two specific large angles (23.48° and 20.32°) of AA-twisted BPC with two specific large angles (23.48° and 20.32°), as shown in Fig. 4(c-d). These flat bands are separated from the other bands and nearly degenerate, which is quite different from the case of TBG or the flatted bands with smaller twisted angles. The band gap of the flat bands is also tunable by varying the twist angle and the modulation range is about 50 meV (detailed information is illustrated in Fig. S4).

To further understand the effect of twist angel on the electronic properties and the underline mechanism of the flat bands, in Figs. 4(e) to (h) we plot the charge density distribution of the flat bands in AA-twisted BPC. It can be found that the charge densities are mainly localized at the AA-stacked region which indicated the flat bands originate from the form of AA-stacked. Meanwhile, as the twist angle decreases, the bandwidth of flat bands is reduced and the local charge density becomes more obvious, which will strongly modify the optical properties and exciton structures of twisted bilayer BPC.

Similar to AA-twisted BPC, we also investigated the electronic properties of AB-twisted BPC. The schematic illustrations of AB-stacked and AB-twisted BPC are plotted in Fig. 5(a-b). The results explain that there is also a phase transition in AB-twisted BPC. It is a metal at 0° (see Fig. 6(a)) while a semiconductor at other angles (only twisted 21.79° (see Fig. 6(b)) is an indirect bandgap semiconductor). In Fig. S5, we show the change of band gap with twist angle for AB-twisted BPC, consistent with AA-twisted BPC, the band gap can also be regulated by twist angle and the variation range of band gap is within 109 meV. Moreover, the flat bands are also observed in the AB-twisted BPC. In addition to the small angles, the structure of twisted 23.48° develops two near-degenerate flat bands near the Fermi level, in which the charge densities are localized at the AA-stacked region, as in AA-twisted BPC, the results depicted in Figs. S7(a) and (d). Different from AA-twisted BPC, when twisted 20.32°, three groups flat bands can be clearly identified near the Fermi level and the number of bands in each group is two (see the labels on the right side of). From Fig. S8(d-e), one can easily see that the charge densities of the first and third groups flat bands are localized at the AA-stacked area, while the second group is localized at the AB-stacked area. More interestingly, two flat bands with energy separation and almost zero bandwidth are generated near the Fermi level when twisted at 16.4° and 14.11° (see Fig. 6(c-d)), and their charge densities are all localized around the AB-stacked area, as shown in Fig. S8 (g-h). It is worth noting that the unit cell with twist angle 16.4° has only 1176 atoms which will be beneficial to the theoretical research of flat bands for twisted bilayer system.

## 4. Conclusion

In summary, we employed the first-principles and tight-binding calculations to study the electronic structures of twisted bilayer BPC. The results show that the direct band gap semiconductor can be converted into metal or indirect band gap semiconductor due to different interlayer interaction. More importantly, the range of band gap can be adjusted from 0meV to 120meV just by changing the twist angle. The band gap ranges from the terahertz to the infrared which is expected to make twisted bilayer BPC a highly promising platforms for light absorption and optoelectronic detection. Furthermore, the production of flat bands in the twisted BPC is independent of the magic angle or the number of atoms. In addition to small angles, flat bands completely separated from other bands can also be produced at several large angles. The charge densities of these flat bands have a variety of local modes, indicating that the properties of these flat bands are different. These results render the twisted bilayer BPC as an ideal and promising candidate to study the physics of strong correlations. Our research also provides a theoretical reference for the twisted bilayer of other 2D materials. We hope these theoretical predictions and findings can be verified by future studies.

# Acknowledgments

This work was supported by the National Natural Science Foundation of China (No. 12074150, No. 11874314).

**Figure Captions**

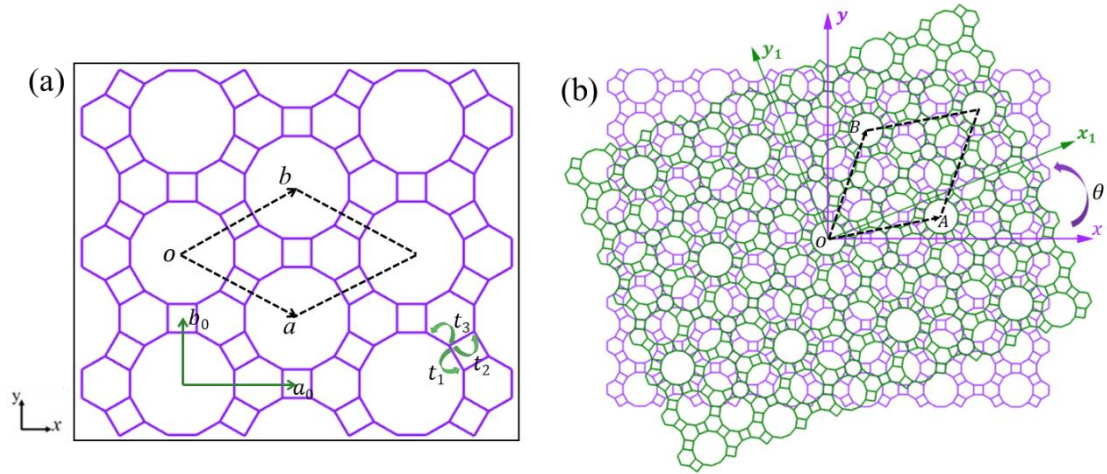

Fig. 1. (a) The structure of AA-stacked BPC. Dashed rhombus indicates the primitive cell for which the lattice constants are a and b, the basis vectors are $\mathbf{a_0}$ and $\mathbf{b_0}$, and the intra-layer hopping parameters are $t_1$, $t_2$ and $t_3$. (b) Schematic illustrations of AA-twisted bilayer BPC system: the green layer is twisted with respect to the purple one underneath by an angle $\theta$ around the common origin (labeled in the figure as *O*). Dashed rhombus indicates the supercell (the primitive cell of AA twisted BPC) for which the lattice constants are *A* and *B*.

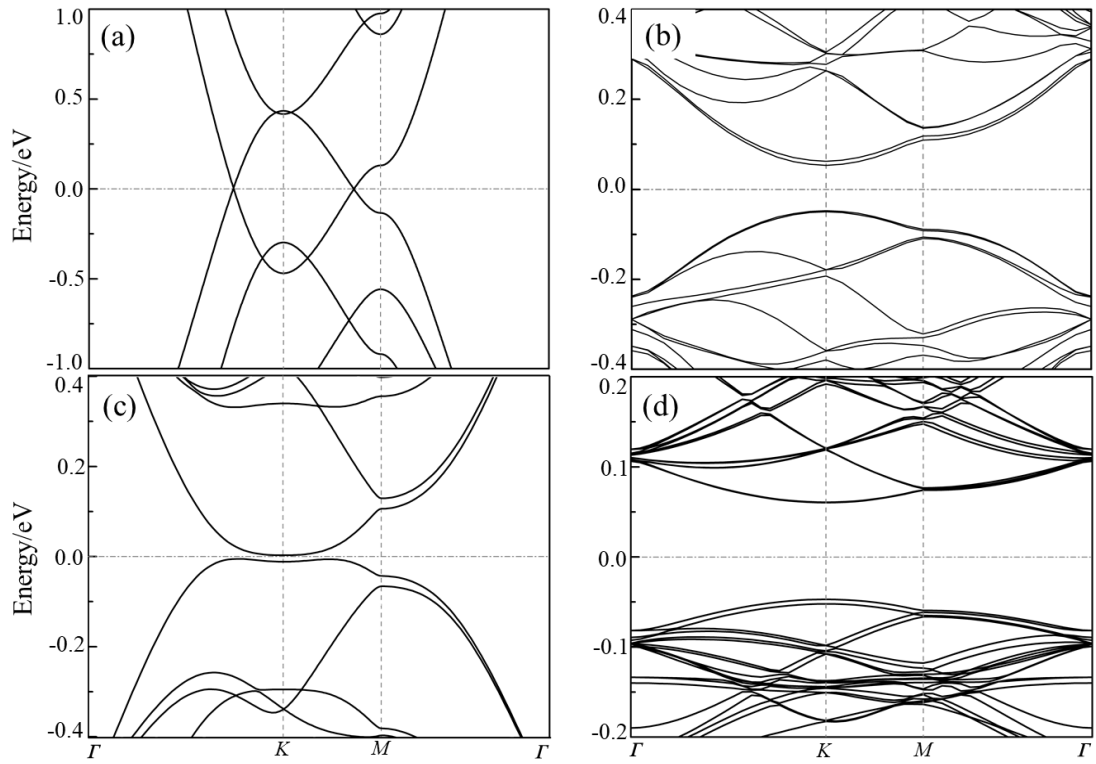

Fig. 2. Band structures of AA-twisted BPC (a) twisted 0°. (b) twisted 13.17°. (c) twisted 21.79°. (d) twisted 29.4°.

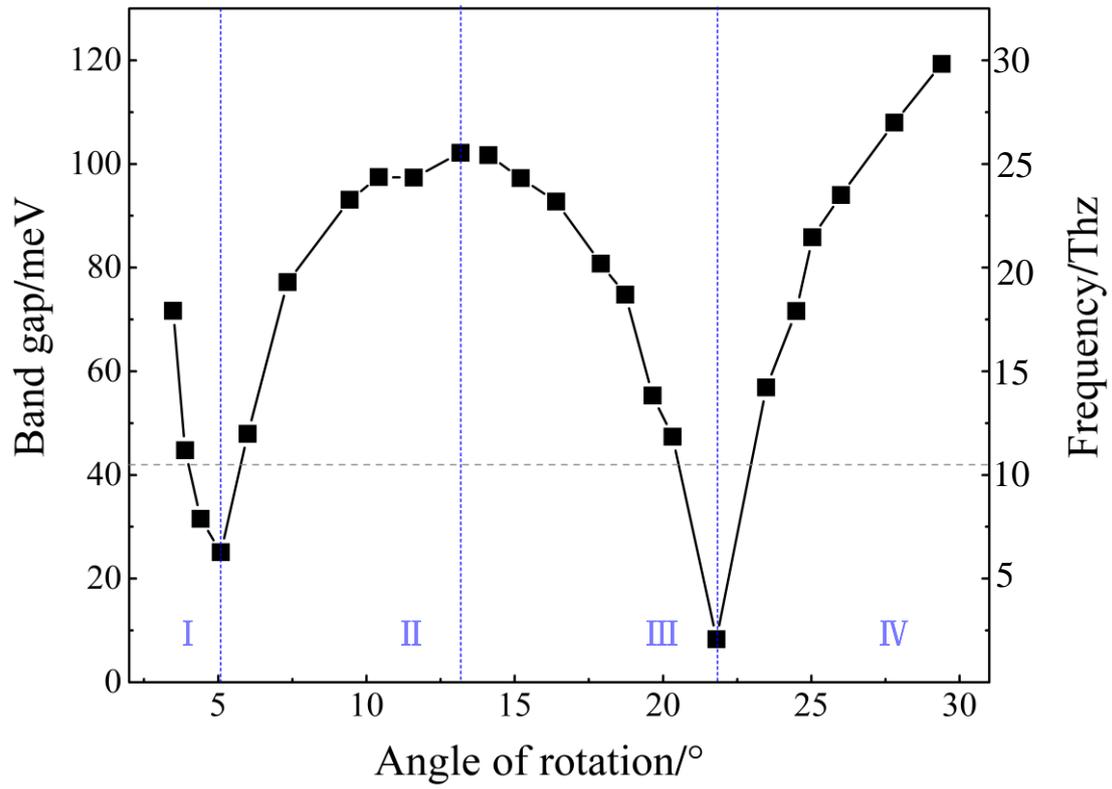

Fig. 3. Band gaps of AA-twisted BPC for different twist angles，gray dotted line is the monolayer BPC result.

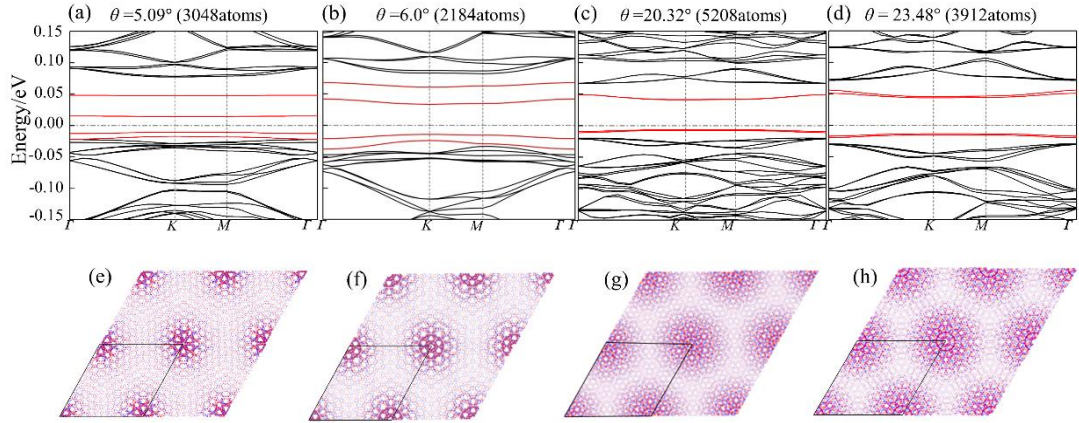

Fig. 4. Band structures of AA-twisted BPC (a) twisted 5.09°. (b) twisted 6.0°. (c) twisted 20.32°. (d) twisted 23.48°. (e-h) Charge densities corresponding to the flat bands in (a-d).

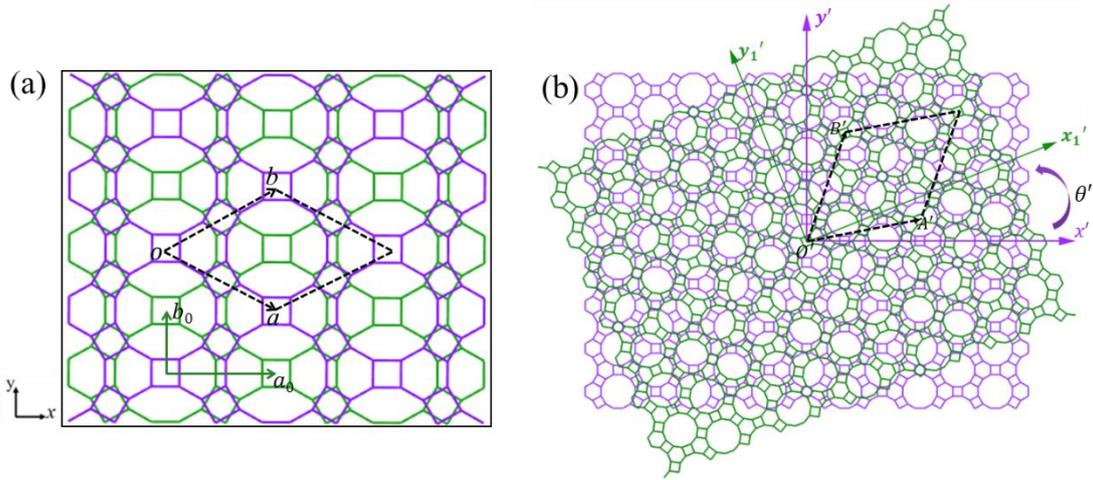

Fig. 5. (a) The structure of AB-stacked BPC. Dashed rhombus indicates the primitive cell for which the lattice constants are a and b, the basis vectors are **a₀** and **b₀**. (b) Schematic illustrations of AB-twisted BPC system: the green layer is twisted with respect to the purple one underneath by an angle $\theta$ around the common origin (labeled in the figure as $o'$). Dashed rhombus indicates the supercell (the primitive cell of AA twisted BPC) for which the lattice constants are $A'$ and $B'$.

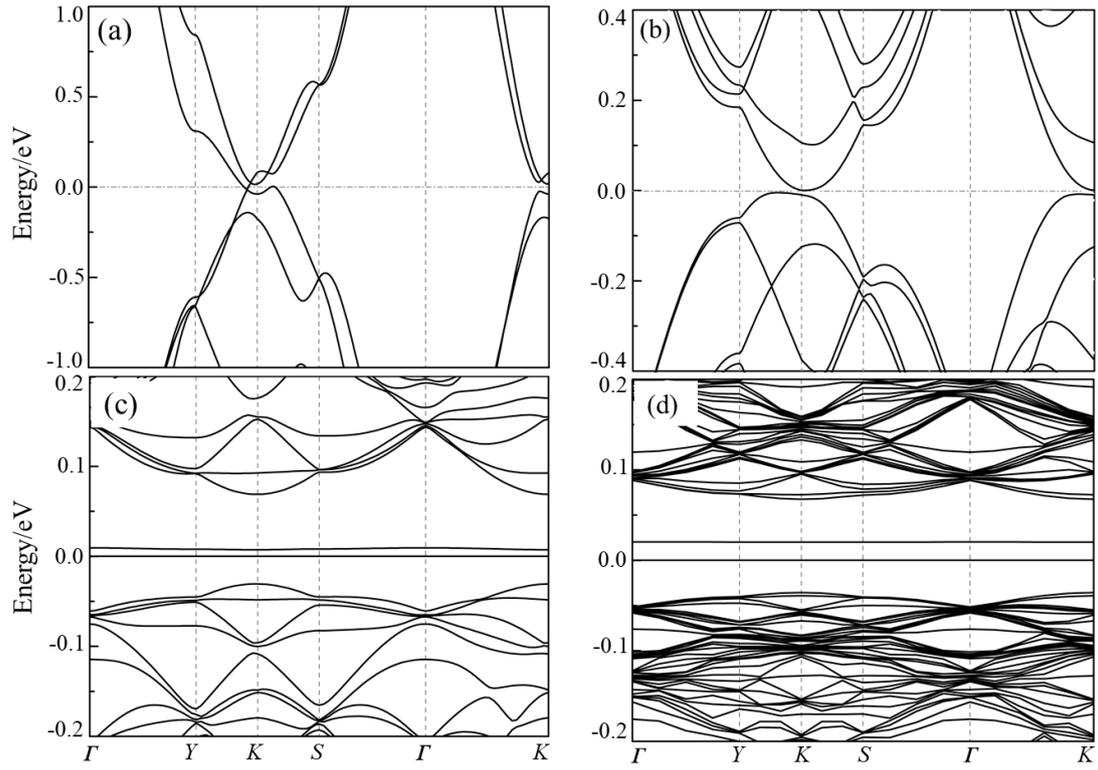

Fig. 6. Band structures of AB-twisted BPC (a) twisted 0°. (b) twisted 21.79°. (c) twisted 16.4°. (d) twisted 14.11°.